**Creating a unique mobile financial services framework for Myanmar: A Review**

Ma Nang Laik*, Chester Mark Hong Wei
Senior lecturer, School of Business
Singapore University of Social Sciences
Singapore

nlma@suss.edu.sg

# Abstract

Myanmar is languishing at the bottom of key international indexes. United Nations considers the country as a structurally weak and vulnerable economy. Yet, from 2011 when Myanmar ended decades of military rule and isolationism and transited towards democracy, its breakneck development has led to many considering the country to be one of the final frontiers for growth in the Asia region. One such industry that has benefitted from the opening of the country is telecommunications. The mobile penetration rate at 4.8% in 2011 has increased significantly to 90% in 2016.

Despite renewed optimism and development in the economy, one statistic remains disappointing. According to a report by Asian Development Bank (ADB), only 23% of the adult population have access to a bank account. This highlights a need to reach out and increase access to financial resources to a population that is severely unbanked and underbanked.

This creates an interesting proposition of allowing both the telecommunications and financial sector to form the mobile financial services (MFS) sector and meet the need of improving access to financial resources for the population. This report explores the government role in supporting, growing and sustaining the MFS sector and conducts a comparative research into Singapore, Malaysia and Thailand to understand the steps taken by these governments to develop their own Financial Technology (FinTech), specifically MFS, industry. Finally, the report will present preliminary recommendations that the Myanmar government could consider implementing to drive growth in its MFS sector.

# 1. Introduction

The Republic of the Union of Myanmar has seen unprecedented progress over the past decade and is widely considered to be one of the final frontiers for growth in the Asian region (Tan, Neo, & Oeni, 2012). Since 2011, Myanmar has undergone a series of political, economic and social changes, started during Thein Sein's leadership and now under National League of Democracy (NLD) (Deloitte, 2013).

One such change is the passing of the Telecommunications Law in 2013, which saw the ownership of SIM cards and internet connectivity shifting from elite's enclave to the lowest-income households in Myanmar (Kean, 2017). The prices of SIM cards fell from US$150-200 in 2010 to

US$2 in 2013-2014 (Nam, Cham, & Halili, 2015), which served to drive up mobile penetration rate. Data from GSMA website places Myanmar's mobile penetration rate at 89.84% and mobile broadband penetration rate at 49.55% as at 2016 (GSMA, 2016). However, a point to note is that the data are calculated using mobile subscription numbers and not unique individuals. Hence, due to the possibility of an individual owning more than 1 mobile subscription, the actual figures may be lower than what was reported on GSMA's connectivity index. Despite these positive statistics, it does not present the lopsided contention between the urban and rural populations. 70% of the population live in rural areas, but the telecommunications infrastructure is biased towards the major cities. This results in poor coverage outside of urban areas, limiting usage accordingly (Deloitte, 2013).

The focus on the rapid growth rate of mobile penetration and broadband is intentional. It highlights the leapfrog nature of Myanmar's society. Majority of the population did not own a mobile phone in 2012 (Deloitte, 2013), yet over the course of 4 years, mobile phone penetration rate not only increased significantly, but smartphones account for more than half of the mobile phones used in Myanmar. This also means that most of the population leapfrogged the usual course of phone ownership and went directly into owning a smartphone from not owning a phone preciously.

Contrasting this, the financial sector in Myanmar remains underdeveloped. The adoption rate of financial services by the population remains low and access to financial services remains challenging. A report published by Asian Development Bank (ADB), in collaboration with Oliver Wyman reported that over a 12-month period, only 23% of Myanmar's adult population have access to a bank account and only 13% save money with a formal financial institution (FI). With the limitation of poor data, ADB estimates that less than 10% of Myanmar's total needs in payments and money transfers are handled by formal FIs, which further supports the challenges shrouding the country's financial sector (Asian Development Bank; Oliver Wyman, 2017). However, Myanmar's financial sector has the potential to succeed. While local banks have been struggling since the liberalisation of the financial sector in 2011, experiencing a decrease of 14% in total banking assets from June 2015 to June 2016, private banks have reported an increase of 27% in total banking assets over the same period (PricewaterhouseCoopers, 2017). Private banks are also experiencing extraordinary growth over the past 6 years, with private banks growing more than 10 times, with total assets growing from approximately US 1.6 billion in March 2010 to USD 17.3 billion in March 2016 (PricewaterhouseCoopers, 2017).

Against the backdrop of robust growth in mobile penetration and limp adoption rate of financial services, it is important to focus on the population demographics of Myanmar. Myanmar has a very young population, with 70% under the age of 40. Moreover, the largest group fall into the "Digital Native" category (Prensky, 2001). This category has been arbitrarily used to described anyone born after 1980, the digital age. Digital natives are comfortable with technology and consider it to be an integral and necessary part of their lives (Prensky, 2001).

In Myanmar, there are two sectors that are positioned for growth soon: Telecommunications and Finance. This also points towards the exciting prospect of a sector that sees the amalgamation of the above-mentioned sectors: mobile banking services. With smartphone penetration and connectivity is increasing rapidly, and a large number of "digital natives" in its population, more opportunities exist in the economy for mobile-based businesses. This drives a focus towards the purpose of this paper, which is the exploration of the potential of a thriving mobile banking industry, and more importantly, how the Burmese government should consider the industry and induct initiatives and or legislation to support the emergence, growth and sustainability of mobile banking in Myanmar.

## 2. Literature Review

The state of FinTech in ASEAN has gained much attention over the past 8 years (United Overseas Bank, 2017). Hence, recent literatures surrounding the industry with a focus on ASEAN have been relatively easy to find. These literatures are also contributed by large organisations, ranging from international financial institutional banks to international organisations such as the United Nations (UN). In addition, I have also referenced to literatures that researched extensively the issue of financial inclusion efforts, with a focus on Myanmar. This is in line with the objective of this project, that the recommendations put forth will seek to support, grow and sustain a mobile financial services environment that will increase access to financial services, especially for the unbanked and underbanked.

A report jointly produced by United Overseas Bank (UOB), Ernst and Young (EY), Singapore FinTech Association (SFA) and ASEAN FinTech Network (AFN) in 2017 focused on the criticality of developing FinTech within ASEAN. This report sets the context of the rise of FinTech on a global scale, seen from *Figure 1*. The FinTech investment activity grew exponentially from 2014 to 2016, rising from US$2.3 billion in 2014 to US$14.4 billion in 2016, representing a 526% increase.

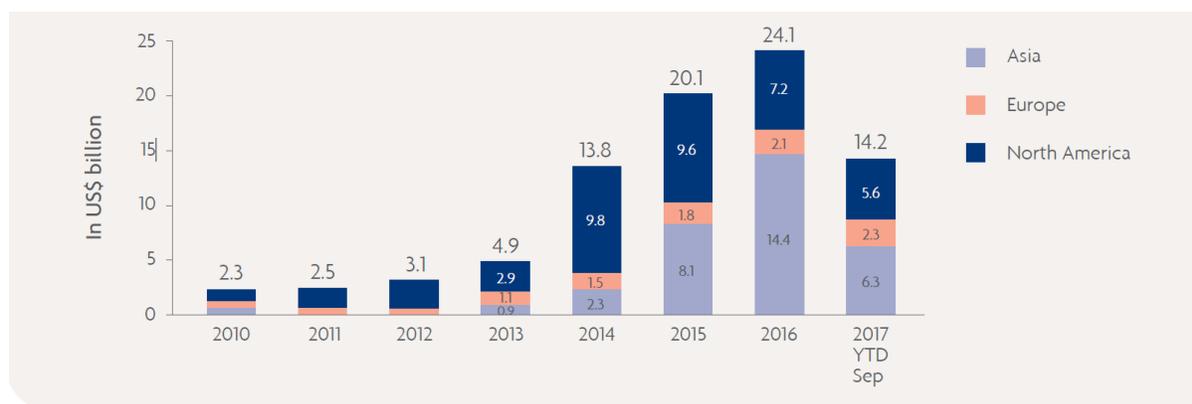

Figure 1: Global FinTech Investment Activity

ASEAN too, witnessed growth in FinTech. This growth is represented both in value of funding received for FinTech products/services/solutions and the volume of funding. Investments in Southeast Asian FinTech market rose 33% from 2015 to 2016 and volume of investment activity during the same period rose nearly 20% (United Overseas Bank, 2017)

The report points out too, that as at 2014, half of the adult population in ASEAN, approximately 264 million adults, are unbanked. The statistic widens in rural areas, with 74% of the rural population being unbanked. So, *ipso facto*, most FinTech initiatives are on payments and mobile wallets to drive greater financial inclusion. Payment and mobile wallet solutions account for 43% of ASEAN's FinTech activities, with financial comparison solutions coming in second at 15%. (United Overseas Bank, 2017). The urgency for FinTech to play a larger role in promoting financial inclusion is even more evident with the low bank penetration in majority of ASEAN countries, as seen in *Figure 2*.

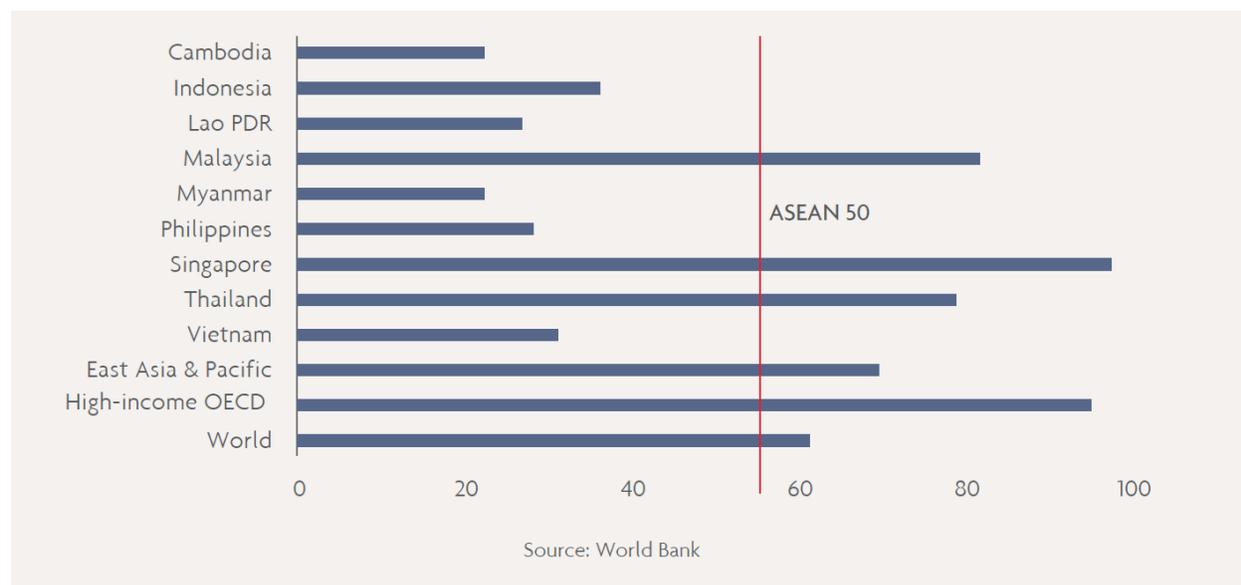

Figure 2: Adults with an account in 2014

Yet these statistics makes it even more attractive and vital for FinTech companies to develop and offer solutions. As mentioned in the introduction of this paper, the adoption of digital mediums is at an all-time high for Myanmar, and this is crucial for the field of FinTech. Myanmar has seen high smartphone penetration rate, and this becomes an indicator of the population's readiness in adopting FinTech, more so with mobile-based financial services.

An extensive research paper was commissioned by the Bill and Melinda Gates Foundation that looked specifically into the opportunities and potential of mobile-based financial services in Myanmar. The research for this paper was conducted with a country-level diagnostic, leveraging data from the Central Bank of Myanmar (CBM), Ministry of Communications, 3 telecommunications companies operating in Myanmar, Facebook, Viber, United Nations Population Fund (UNFPA) and a representative Information and Communication Technology survey of 7,500 households. The paper holds that Myanmar provides an ideal lab environment to develop an entire ecosystem of mobile-based financial services that could not only disrupt the cycle of poverty but also become an integral part of everyday lives. This literature also touched briefly on the impacts of governmental policies and collaboration with other governments and organisations. It pointed out the importance of influencing and shaping the mobile financial services industry, to accelerate the implementation, adoption and impact across the ecosystem. Crucially, all these need to be done to push the industry to reach all parts of Myanmar, even the poorest and most vulnerable of areas (Htun & Bock, 2017).

Tracing the existing literatures, ASEAN is poised for exponential growth in FinTech. This is unsurprising considering the rise of FinTech globally. Extrapolating the rate of growth within

ASEAN, it would point towards Myanmar partaking in this growth. The report commissioned by Bill and Belinda Gates Foundation investigated the potential of mobile-based financial services, of which this will be the focal point of interest in this paper. Yet, all existing literatures hardly explores the relationship between regulation and encouraging FinTech activity. Moreover, there is a wealth of knowledge that has been accumulated by the experiences of other jurisdictions' efforts to incubate, grow and sustain FinTech activities, while at the same time, ensuring that there are appropriate regulations and or safeguards that protects not only the consumers but the larger financial ecosystem as well. As such, the methodology taken in this report will be a comparative analysis of 3 ASEAN countries, namely Singapore, Malaysia and Thailand. This is done to understand what these macrosocial units have done specifically to encourage FinTech activity and the steps they have taken to ensure that the industry grows sustainably and in a manner that does not impose unreasonable risk on the entire economy. Finally, these findings will lead to preliminary recommendations for Myanmar – understanding best practices around the region, learning from their neighbours' experiences and localizing the findings for potential implementation by the government.

## 3.0 Methodology

It has been highlighted in the earlier paragraphs that both the telecommunications sector and financial services sector are poised for growth, and this will also lead to an exciting prospect for a mobile financial services sector to emerge from the amalgamation of both sectors. Yet, the risk of not regulating this new sector may undermine the benefits, especially long-term, that mobile financial services might bring to the economy. Additionally, it is important not to over-regulate the sector as well, to prevent stifling growth in the sector through the setting of too many rules and over-bureaucracy.

This paper provides a comparative analysis of governmental frameworks and policies across 3 countries, namely Singapore, Malaysia and Thailand. According to *Figure 3* below, these are selected to their popularity in ASEAN for FinTech companies which serves as evidences for good infrastructure, supportive policies/ regulations, mobile penetration, capable workforce (United Overseas Bank, 2017) and the relatively closer cultural similarities. The main objective for doing so is explained in the above section - to develop a unique framework to govern the mobile banking sector for Myanmar.

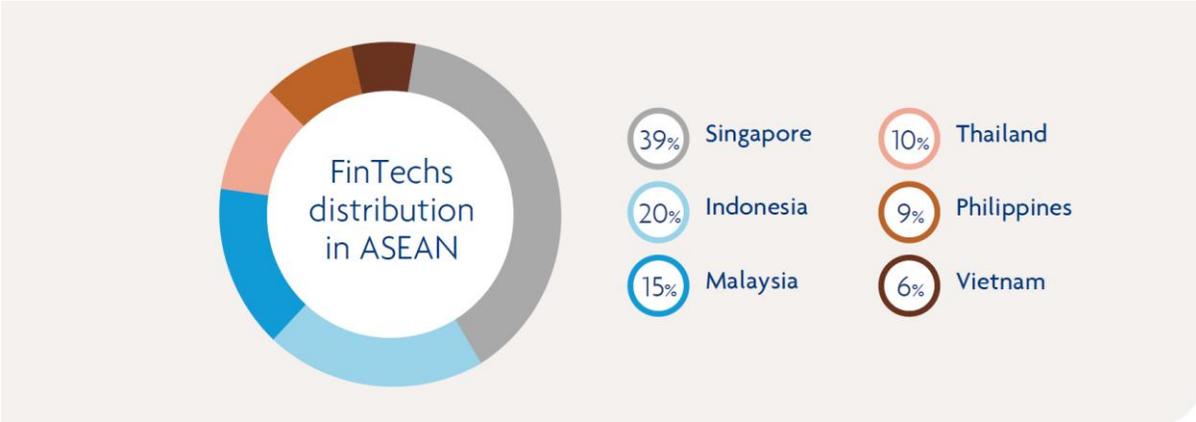

Figure 3: Distribution of FinTech companies among ASEAN

Comparison is inherent in all echelons of a society, and a research methodology based off this is fundamental. Generally, carrying out a comparative analysis would lead to an identification process of the similarities and differences between macrosocial units - countries, nations and other larger political entities. The importance of drawing such information lies in understanding and interpreting these processes and the significance of such knowledge (Ragin, 1987).

Using a comparative method in the discipline of political science has a standardized definition referring to the systematic analysis of a small number of cases, otherwise known as a "small N" (Collier, 1993). There have been many literatures detailing the comparative method, and these formed a basic understanding of the practice of "small N" studies. One of the most important and outstanding literature is from Arend Lijphart's 1971 article titles "Comparative Politics and Comparative Method." His work found that data collection using the comparative method demanded less resources as compared to experimental or statistical methods. Hence, it is most appropriate in researches that face scarcity of time, energy and financial resources because an intensive comparative analysis with a lack of resources may prove to be more promising than a more superficial analysis of many cases. And in such a situation, it will be fruitful to consider the comparative analysis method as the first step of research, setting the stage for future statistical analysis to be carried out on the back of this research (Lijphart, 1971).

Yet, there are some opponents towards this methodology - citing resultant biases when comparisons are made across political and social systems that are territorial (Pennings *et al*, 1999) and also the advantages and disadvantages of selecting countries as 'comparators', with one such disadvantage being the obscurity of domestic differences of a particular country when being compared to international differences and or standards (Arnove, Kelly, & Altbach, 1982).

We will be applying the comparative analysis across territorially distinct macrosocial units as explained in my previous paragraphs and will also adopt the viewpoint that the countries selected are distinct enough to provide valuable knowledge. In addition, we will also attempt to localize the information into relevant material for the context of Myanmar.

The justification for using comparative analysis as the governing methodology for this paper are:
- To identify similarities and differences arising from the practices, processes and policies from the various jurisdictions
- To understand the conditions of which the frameworks and policies had been created from
- To study the impact of governmental support through policies on the FinTech industry, with a focus on mobile financial services
- The scarcity of time, energy and financial resources

In this paper, we propose a theoretical roadmap for the Myanmar government in building a conducive environment for mobile financial services as an industry to thrive. This proposed roadmap is done predominantly by comparing and analysing other countries' governmental initiatives and policies towards this industry. The countries selected based on their efforts and success in encouraging growth in the FinTech industry are, Singapore, Malaysia and Thailand. These countries are part of ASEAN and selected based on the progress made by these governments and the proximity and closer cultural similarities to Myanmar.

## 4.0 Preliminary Findings

### 4.1 Singapore's FinTech Space

#### 4.1.1 Smart Nation Drive

Singapore has enjoyed success in its push towards transforming into a smart nation. Since Singapore's Prime Minister, Mr Lee Hsien Loong, introduced this focus (Lee, 2014), the Smart Nation Programme Office has been set up to coordinate efforts and initiatives across all Government departments to fulfil the mission of building a Smart Nation that effectively benefits everyone. Since then till now, Singapore has climbed the ranks to be one of the forerunners of smart cities globally[1].

Singapore laid out 3 key pillars in its pursuit of its Smart Nation goals. They are Digital Economy Framework for Action, Digital Government Blueprint and Digital Readiness Blueprint. The Digital Economy Framework for Action is focused on encouraging Singapore to thrive in the digital economy. This framework spells out the Government's strategy to guide collective efforts in keeping Singapore ahead of the ever-changing digital economy landscape. This includes infrastructure spending, other investments to increase Singapore technological capacity and capabilities, to name a few. The other 2 pillars, which are not as relevant, shows the Government's efforts to transform public service by leveraging data and using new technologies and to increase digital literacy among its citizenry, ensuring that the population are using technology responsibly and serves to benefit themselves and the society, respectively.

In addition, to drive adoption of technologies, Key Strategic National Projects have been identified as key enablers in Singapore's Smart Nation drive. Out of these, there is a focus on e-Payments which aims to allow everyone to make payments in a simple, swift, seamless, and safe manner. For the purposes of this report, the focus will be on the Digital Economy Framework for Action and e-Payments.

#### 4.1.2 Digital Economy Framework for Action

The Singapore Government intends to equip businesses, workers and the populace to harness the potential and advantages of the digital economy. This is done through 3 pillars, Accelerate, Compete and Transform. These pillars will be supported by 4 enablers, Talent; Research and Innovation; Policy, Regulations and Standards; Physical and Digital Infrastructure.

Accelerate: To increase the pace of digitalisation of existing industries for productivity improvements, efficiency gains and new revenue opportunities.

Compete: To increase economic competitiveness by fostering new integrated ecosystems centred around customers' needs

Transform: To develop the next-gen digital industry as growth engine for the economy and drive digitalisation across industries

With a keen eye on the usage of mobile technologies to effect change on the wider digital economy, Singapore has invested heavily in its physical and digital infrastructure on which future digitalised products and services have a strong foundation to be developed upon, and users have the necessary infrastructure to adopt mobile technologies. Singapore comes in first in terms of 4G speed by OpenSignal and sees a wireless broadband penetration rate of 200% and a mobile phone penetration rate of 150%[2].

---

[1] https://www.businesstimes.com.sg/government-economy/singapore-tops-global-smart-city-performance-ranking-in-2017-study, https://www.channelnewsasia.com/news/singapore/singapore-best-performing-smart-city-globally-study-10038722, https://www.straitstimes.com/business/spore-pips-london-ny-to-top-global-smart-city-ranking

In addition to infrastructure spending that empowers innovation, the report outlined a S$225 million fund that has been created to support FinTech efforts that will contribute to Singapore's push to become a digital financial centre. As such, Singapore currently ranks 1st in the 2018 Institute for Financial Services Zug (IFZ) Global FinTech Rankings.

### 4.1.3 E-payments

The Singapore Government has made clear its focus on developing the E-payments space. The view that e-payments can

1) make transactions simple, swift, and safe for both consumers and businesses
2) boost digitisation of business processes, enhance productivity, reduce costs and create new business models in the digital economy

provides justification to invest into E-payments (Smart Nation and Digital Goverment Office; Monetary Authority of Singapore; The Association of Banks in Singapore, 2017). To run this, the Monetary Authority of Singapore (MAS) has been tasked spearhead this and to determine the right approach and or regulations towards this area. MAS functions as Singapore's central bank. Its role includes formulating monetary policy and identifying emerging trends and potential vulnerabilities within the financial industry. It also operates on a mandate which seeks to foster a sound and progressive financial services sector in Singapore. This is achieved by enabling a close working relationship between government agencies and FIs.

The last point on the importance of public-private cooperation is evident. The Institute for Infocomm Research, a government agency, serves as a platform to connect FIs with scientists who are seeking to commercialise their research.

Moving from this, with the nation's focus on building a Smart Nation with transforming the digital economy as a pillar and e-Payments as a Key Strategic National Project, the focus now shifts towards MAS and their role in regulating this space, specifically the focus of this paper – FinTech. The next 2 sections will look into MAS initiatives, namely the regulatory sandbox and the Payment Services Bill.

### 4.1.4 Regulatory Sandbox

MAS approach towards the burgeoning FinTech industry is not to create regulations. On the contrary, MAS undertook a "regulatory sandbox" approach. This approach allows banks to experiment their FinTech solutions without the need to obtain regulatory approval in advance (Vasagar & Weinland, 2016).

In June 2016, MAS created FinTech Regulatory Sandbox Guidelines which sets up to encourage more FinTech experimentation so that innovations can be tested in the market and have a chance for wider adoption. These experimentations will be carried out within a well-defined scope which includes duration and space, and the sandbox will include safeguards to contain any repercussions or consequences of failures so that the overall financial system continues to be safe. MAS also maintains the right over the legal and regulatory requirements which will either be tightened, maintained or relaxed for every experiment. The sandbox model aims to seek a "sweet-spot" in the trade-off between regulation and innovation, by subjecting FinTech entities to regulations upon proving that the solution put forward has 1) achieved its intended outcomes in test scenarios and 2) that it will be able to satisfy and be compliant to the relevant legal and regulatory requirements.

---

[2] https://www.imda.gov.sg/industry-development/facts-and-figures/telecommunications#1x

The steps needed by a company to enter the regulatory sandbox is relatively straightforward, as seen in *Figure 4*:

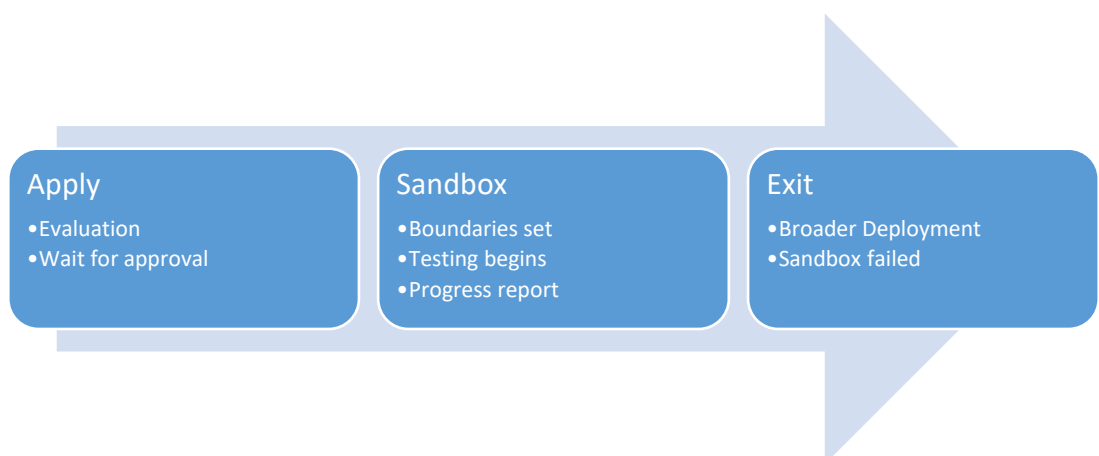

Figure 4 Process of applying and participating in the Regulatory Sandbox

While *Figure 4* is simplified to explain the process for an entity to apply and go through the regulatory sandbox, there are criteria that entities must adhere to. MAS developed 2 sections in its guidelines that serves as an eligibility check for entities. Only when an entity is confident that its products/services meet the objective, principles and criteria set out in these sections, then they are to send in their application. Although MAS casts a wide net to encourage applicants to participate in the sandbox, the sections developed act as an invisible barrier for entities looking to apply.

This invisible barrier provides some order in sifting out applications from the onset. There are regulations that are to be maintained, such as adherence to the Personal Data Protection Act (PDPA), Anti-Money Laundering (AML)/ Countering the Financing of Terrorism (CFT) regulations. MAS is lax on the wording of the application requirements, which is summed up in paragraph 2.3 of the FinTech Regulatory Sandbox Guidelines as "Depending on the financial service to be experimented, the applicant involved, and the application made, MAS will determine the specific legal and regulatory requirements which it is prepared to relax for each case." MAS provides examples of these legal and regulatory requirements in Annex A of the guidelines, and while examples, the requirements that are to be maintained involves PDPA and AML/CFT.

However, FIs are already free to launch new financial products, services or processes without first seeking MAS' permission (MAS, 2016), and with this, it is probable that new FinTech may be introduced through it. This begs many questions over the effectiveness of the sandbox model, and who are the target audience for the sandbox.

Overall, the intent is good - to encourage innovation and provide a safe environment to test promising innovations.

### 4.1.5 Payment Services Bill

Payment services in Singapore have been regulated under 2 separate legislations, namely the Payment Systems (Oversight) Act (Cap. 222A) and the Money-Changing and Remittance Businesses Act (Cap. 187). However, MAS recognises that the landscape for payment services has changed significantly over the years and is proposing to:

1) streamline the 2 legislations above

2) enhance the scope of regulated activities

3) adopt a modular regulatory approach to calibrate regulations according the risks posed by the relevant activities

These considerations give rise to the proposed Payment Services (PS) bill, with a key intent to promote greater confidence among consumers and merchants to adopt e-payments. This is in line with the nation's push into e-payments as a focus area.

A report issued by PwC titled "Strengthening Singapore's payment services through regulation" provides a succinct overview of the proposed PS bill and the intended impact it brings. This bill signals MAS' departure from product-based licensing model into an activities-based and risk-based licensing model. This is significant. Having a "blanket" license with the previous approach fails to consider new payment services model, bringing about risks in the economy. Hence, having a modular activities-based and risk-based licensing model would allow MAS to expand the scope of regulated activities and implement risk-mitigating measures towards licensees' activities, wherever relevant. In this modular approach, payment institutions will see their activities subjected to approval, but the activities are not licensed individually (PwC, 2018). So, all retail payment services providers will be classified under one of the following categories:

- Major Payment Institutions (this includes operators of existing widely accepted stored-value facilities)
- Standard Payment Institutions
- Money-Changers – which can only provide money-changing services

In my view, I think this is necessary and good for payment services providers as it reduces the regulatory and bureaucratic burden to comply.

On the risk-based licensing model, the PS bill outlines 4 key risks that are applicable to payment services providers

- Money laundering/ terrorism financing (ML/TF)
  Using online payment methods as a medium for ML/TF
- User protection
  Transaction disputes and insolvency of e-money issuers are adequately addressed
- Interoperability

> There must be 3rd-party access to payment systems on fair and reasonable commercial terms. It is also mandatory to participate in a common platform and or an interoperable common standard for payments, such as a standardised QR code.
> - Technology risk
> Ensure cybersecurity through user authentication, prevention of data loss and fraud monitoring and or protection

The ML/TF requirement will apply to all the 3 licenses listed above, whereas the remaining requirements will apply to payment institutions only.

In the PS bill, there are proposed exclusions in 2 categories:

1) Carving out of 3 specific activities, namely limited purpose e-money and virtual currency, and incidental or necessary payment activities. These include loyalty programmes or public authority pre-paid cards. Limited purpose virtual currencies refer to in-game credits and are to be distinguished from Bitcoin and Ether.
2) Specific application of risk-mitigation measure. ML/TF mitigations will not apply to certain low risk services such as money transfer services used for payment for goods or services from an identifiable source. However, MAS is seeking feedback on whether payment for goods on e-commerce platforms should be classified as low risk.

In this report, PwC is of the view that the proposed changes in the PS bill will become a catalyst for the development of Singapore's payments industry and landscape, enhancing its competitive edge. However, MAS must be thoughtful and sensitive in the implementation of this bill. The bill addresses some questions regarding the transitional period during implementation. Entities that are considered low risk would qualify for specific exemptions. A 6-month grace period will be extended to in-scope entities that are currently not subject to regulatory requirements. This grace period will allow entities time to be compliant to regulations. Due to the fragmented nature of payment services, there will be a deferred start for interoperability requirements, again allowing time to providers to introduce such measures into their payment services. Lastly, there will be transitional provisions for regulated FIs and payment firms (PwC, 2018). While the proposed bill is short of details, these arrangements to ease the transition from PS(O)A and MCRBA should be received positively by providers.

## 4.2 Malaysia's FinTech Space

In Bank Negara Malaysia's (BNM) 10-year financial blueprint issued in 2011, it identified the strategic importance of using e-payments as one of the 9 focus areas to achieve the main goal of transforming Malaysia into a and high-income and value-added economy. BNM has recognised the need to collaborate with all payment system stakeholders to improve the entire payment process and change payment behaviour. Hence, the Bank has committed to provide fair and healthy competition in the retail payments space. However, the Bank acknowledges that achieving economies of scale is of equal importance to generate cost efficiency. In its blueprint, BNM believe economies of scale and cost efficiency can be achieved when there is one or a small number of players that exist in the sector (Bank Negara Malaysia, 2011). This brings about the need to achieve a fine balance between fair competition and having one major or a small number of players in the sector. BNM have decided that having close oversight and regulatory prescriptions to prevent anti-competitive behaviour would allow this fine balance to be achieved.

BNM laid out markers to work towards and has committed to accelerate migration into e-payments. These markers can be seen in the table below.

|  | 2018 figures | Target by 2020 |
|---|---|---|
| E-payment transactions per capita | 111 | 200 |
| Debit card transactions per capita | 5.1 (2017 figures) | 30 |
| No. of electronic funds transfer at point of sale (EFTPOS) terminal per 1,000 inhabitants | 14 | 25 |
| Number of cheques cleared | 119 million | 100 million |

Table *1*: Extracted from Bank Negara Malaysia

BNM's focus on e-payments come with the viewpoint that e-payments would allow businesses and the larger society to enjoy convenience and operational efficiency that would come from the expedited nature in payments and receiving funds. A speech by the Governor of BNM, Muhammad bin Ibrahim in 2018 seems to present that Malaysia is on track to meeting the markers listed above and societal changes that fit into the macro-level benefits. For ease of reading, I have input the 2018 markers into Table 1. As seen, it seems unlikely that BNM would achieve its targets. This was evident in the same speech, where the Governor alluded that cash-based transactions remains prevalent among retail transactions in the country, especially with lower tier merchants, who prefer dealing with cash.

Unlike Singapore, Malaysia does not have a macro policy objective of digital transformation. Instead, BNM has identified e-Payments to be an area of strategic importance and has devoted resources to encourage consumers to utilise this mode of payment. Hence, the next section will investigate BNM's efforts in the e-Payments space before analysing the regulations and safeguards proposed by BNM, namely the regulatory sandbox.

### 4.2.1 Mobile payments and e-wallet systems

In another speech by the Governor of BNM in 2017, he provided more specifics on how the medium of mobile phones should be optimised for payments. This is more relevant to the aim of this report. BNM opines that mobile payments can become a viable, low-cost alternative to using cash and complement card payments. To push the envelope further, e-wallet systems are viable as merchants can accept payment through a QR code. It will be cheaper for consumers to use QR code as opposed to paying the Merchant Discount Rate (MDR) that is levied on card payments. The usage of QR code is significant because it enables financial inclusion. In Malaysia, there are 24 million adults and of this, 10 million are not using online services and 2 million are unbanked. Having non-bank e-payment providers would serve this segments, and QR code can become the catalyst for this service. To this end, BNM has introduced an Interoperable Credit Transfer Framework (ICTF) to achieve 2 objectives, namely avoid market fragmentation and broaden financial inclusion. The ICTF will link both banks and non-bank e-money providers to ensure that consumers are able to reach their bank accounts and or e-money accounts. This way, it will allow all consumers to transfer funds seamlessly by mobile numbers, identity card numbers or scanning the QR code. BNM has identified PayNet as the operator of Malaysia's shared payment infrastructure. This infrastructure will provide open and unbiased access to both banks and non-bank e-money issuers. Having a common infrastructure would shift the focus of competition among market players away from infrastructure and towards value-added services (Muhammad, 2017).

### 4.2.2 Financial Technology Enabler Group (FTEG)

The FTEG was started by BNM in 2016 to support FinTech innovations that increases quality, efficiency and access to financial services in Malaysia. They are responsible for formulating and enhancing regulatory policies to facilitate and accelerate, wherever applicable, the adoption of FinTech. The FTEG is responsible, together with BNM, for the launch of Malaysia's FinTech regulatory sandbox, which will be expounded in the following section.

### 4.2.3 Regulatory Sandbox

The launch of the fintech regulatory sandbox is needed given the rise of fintech solutions and innovative business models. The implications of this has led to BNM seeking to provide a regulatory environment that is conducive for FinTech deployment – which includes the review and adaptation of regulations or procedures that are anti-innovation. Implied in this, the sandbox is used to catch regulatory and compliance policies that may inhibit innovation by looking at the operations of FinTech start-ups and their products/services (Reyes, 2017). A sandbox enables innovative FinTech to be deployed and tested in a live environment, given specified parameters and timeframe. However, with the sandbox operating in a live environment, risks persist and may cause financial loss to sandbox participants and or customers. Thus, BNM recognises the importance of incorporating safeguards in the sandbox model for risk management purposes in the event of failures. This will be further explained later. In addition, the sandbox cannot be used by participants seeking to get around existing laws and regulations. So, it is not suitable for any proposals that are already addressed under prevailing regulations – if the proposals fall into this, BNM will take an "Informal Steer" approach and provide guidance and advice on the modifications needed on the proposals to align it to prevailing laws. Hence, these form the principles on the necessity and importance of having a regulatory sandbox.

The objective, then, for the regulatory sandbox is to encourage deployment of fintech in a regulatory environment, essentially finding the balance between regulation and innovation. BNM will factor in 3 main considerations for the extent of flexibilities to exercise on regulations for sandbox participants:

1) Potential benefits of the proposed product/service/solution
2) Potential risks and respective mitigating measures
3) Integrity, capability and track record of FIs or fintech companies

BNM outlined the eligibility criteria in a section. The approval to participate in the sandbox can be summarised into the points below:

- The proposed product/service/solution is innovative with potential to improve the following factors:
    - Accessibility, efficiency, security and quality of financial services
    - Efficiency and effectiveness of risk management
    - Existing gaps or open new opportunities on financing and investments

- Applicant has assessed that the product is useful and functional, and has identified the relevant risks
- Applicant has sufficient resources to support the sandbox testing, which includes buffer and expertise to mitigate potential risks and losses during the test period
- There must be a realistic business plan for deployment on a commercial scale upon exit from the sandbox
- Proposed product/service/solution must not be governed under prevailing laws
- Applicant must have a management team or board of directors with proven track record

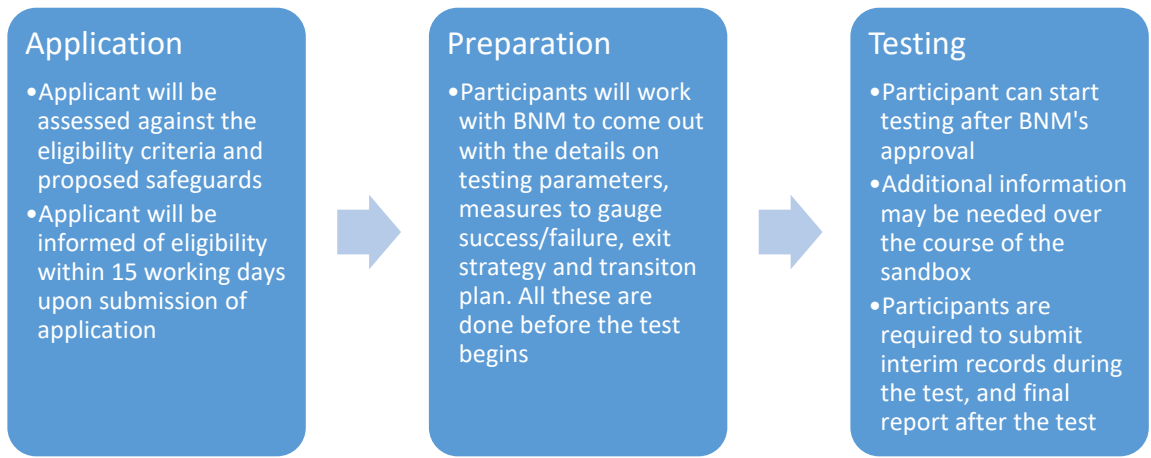

Figure 5 Process of applying and participating in the Regulatory Sandbox*

*Note that for all rejected applications, there will be 6 months cooling off period before the applicant can resubmit the application

In addition, advantages are afforded to fintech companies that collaborate with financial institutions, especially in guidance and resources offered by FIs. The advantages spelled out in the framework are in regulatory requirements and risk mitigations. Also, fintech companies with the potential to contribute to the creation of high value-added jobs will be more favourably assessed.

In its next section, BNM has placed the onus on the applicant to identify the potential risks that its proposed solution may have towards the FIs and consumers during the testing period. After doing so, the applicant too must propose appropriate safeguards to address these risks. This action is in line with the second point of the eligibility criteria listed above. BNM will assess the risks and proposed safeguards according to the following points:

**Risk**

- Preserves sound financial and business practices that are consistent with financial stability
- Treats consumers fairly
- Prevents ML/TF
- Ensures customers' information are kept confidential

- Payment systems and instruments must be safe, reliable and efficient
- Innovative Islamic financial services must be compliant to Shariah standards
- Healthy competition among financial products and services

**Safeguards**

- Adequate disclosure of potential risks to consumers during the testing period, of which consumers must show understand and acceptance of risks
- Limit the number of consumers and or the aggregate value or frequency of transactions during the testing period
- Limit consumers' participation to a certain segment or consumer's profile
- Limit duration of the testing period
- Provide an avenue for consumer redress, including the option for financial compensation that is claimable under specified circumstances
- Committing adequate and competent resources, both financial and expertise, to lead and manage the testing. This is needed for the implementation of risk mitigation solutions in the event of failure.

However, we have some counter arguments over the role of BNM in the regulatory sandbox and the purpose of the sandbox. There are a couple of points raised in the above paragraphs that seem to be conflicting with one another. They are:

1. One objective of the sandbox is to catch regulatory and compliance policies that may inhibit innovation by looking at the operations of FinTech start-ups and their products/services
2. The proposed product/service/solution is genuinely innovative
3. Advantages are afforded to fintech companies that collaborate with financial institutions, especially in guidance and resources offered by FIs
4. BNM has placed the onus on the applicant to identify the potential risks that its proposed solution may have towards the FIs and consumers during the testing period

However, this objective seems to confuse 1) the role of BNM and 2) the purpose of the sandbox.

1) The role of BNM

    Basing this from the second, third and fourth point, it would be difficult to pinpoint BNM's role during the sandbox testing period. In the second point, the wording used is genuinely innovative. This seems to suggest that BNM are in favour of a proposed solution being new in the market. Yet, the framework does not qualify what constitutes a solution as new. However, the greater point is limiting this at the sandbox stage would limit the element of innovation, which I feel is crucial in the financial ecosystem to ensure that entities are always improving themselves and seek to serve the larger economy.

    In the third point, affording advantages to fintech companies that collaborate with FIs seems to be conflicting with the spirit of innovation. Small fintech companies could have breakthrough technologies but lack the accessibility to funding or to institutions. Hence in applying for sandbox testing, they may be given up for a fintech company that has collaborated with an FI. However, BNM could act as a catalyst for this, linking up their

network of FIs to such fintech companies should the proposed solution meet the eligibility criteria spelled out in the framework

In the fourth point, would not the role of BNM be to identify the potential risks of the proposed solution and apply appropriate safeguards? The onus seems to fall squarely on the applicant. For BNM to effectively carry out its role as a regulator, it would be important to be in the process, with the applicant, to identify risks and the relevant mitigation measures.

BNM should assess applicants based on the potential of the proposed solution and the level of innovation in the underlying technology. While this assessment would factor in risk, BNM role as a regulator would be to provide clear guidelines on the severity of different types of risks, and to provide guidance and oversight to manage such risks. In all, I feel that there is a need to clearly distinguish the dual roles of assessment and regulator – this could take on the form of creating different teams to assess and manage risk or to establish a different set of protocols in the application process.

2) Purpose of the sandbox

   The sandbox seems to serve a Proof-Of-Concept (POC) model to BNM with the abovementioned objective. That should not be part of or the purpose of the sandbox – educating BNM on new, potentially proprietary financial technology. I have 3 arguments to support my view.

   First, sandbox is not the medium for education. The main aim for companies to utilise the sandbox is to test their solutions on a small group of consumers with the view on widespread adoption upon exit of the sandbox. Instead, if it is used for education, it would divert the focus of these companies away from their users and towards regulators. This could increase compliance burdens for the companies which may prove to be costly.

   Second, the erosion of advantage through observation. Before listing my argument, the framework stated that for products, services or solutions that are not suitable to be in the sandbox, BNM will take the approach of an "Informal Steer". This is done to provide guidance and advice to FIs or fintech companies on the modifications they can do to align their business models or solutions with prevailing laws and regulations (Bank Negara Malaysia, 2016). This consultative approach is problematic. Competitive advantages for companies are usually established through its operations and or products/services offered. With BNM observing companies in the sandbox, the company's technology and processes will be made available. These can turn out to be sensitive. Linking it back to "informal steer", questions have to be asked on how sandbox participants' business processes and technology can remain confidential as BNM provide guidance and advice to other FIs or fintech companies outside of the sandbox.

   Third, the strain on resources. Having the sandbox as a POC model would place a strain on resources for early entrants. While the regulators are being educated of new FinTech solutions and processes, concurrently they must be able to identify if the proposed solution shows the potential to effect positive, value-added impact to the larger economy. The hypothesis here is for the earlier entrants, this process of evaluating the potential proposed solution would be less efficient, causing it to be dragged on. For example, the regulator will take longer to decide that a fintech start-up's product would not bring positive change to the economy and waste the start-up's resources for an extended period. Comparing this to a similar product that reappears in the sandbox after 6 months, the regulator would be quicker to issue a "No". In all of these, by having the sandbox operate as a POC, it may benefit the regulators, but it would not benefit start-ups or companies that have finite resources.

## 4.3 Thailand FinTech Space

### 4.3.1 Thailand 4.0

In 2016, Thai's military junta unveiled its policy named "Thailand 4.0". It is an industrial policy that is sector-specific aiming to transform the economy. The transformation will be defined by innovative technology-based manufacturing and services (The Economist, 2017).

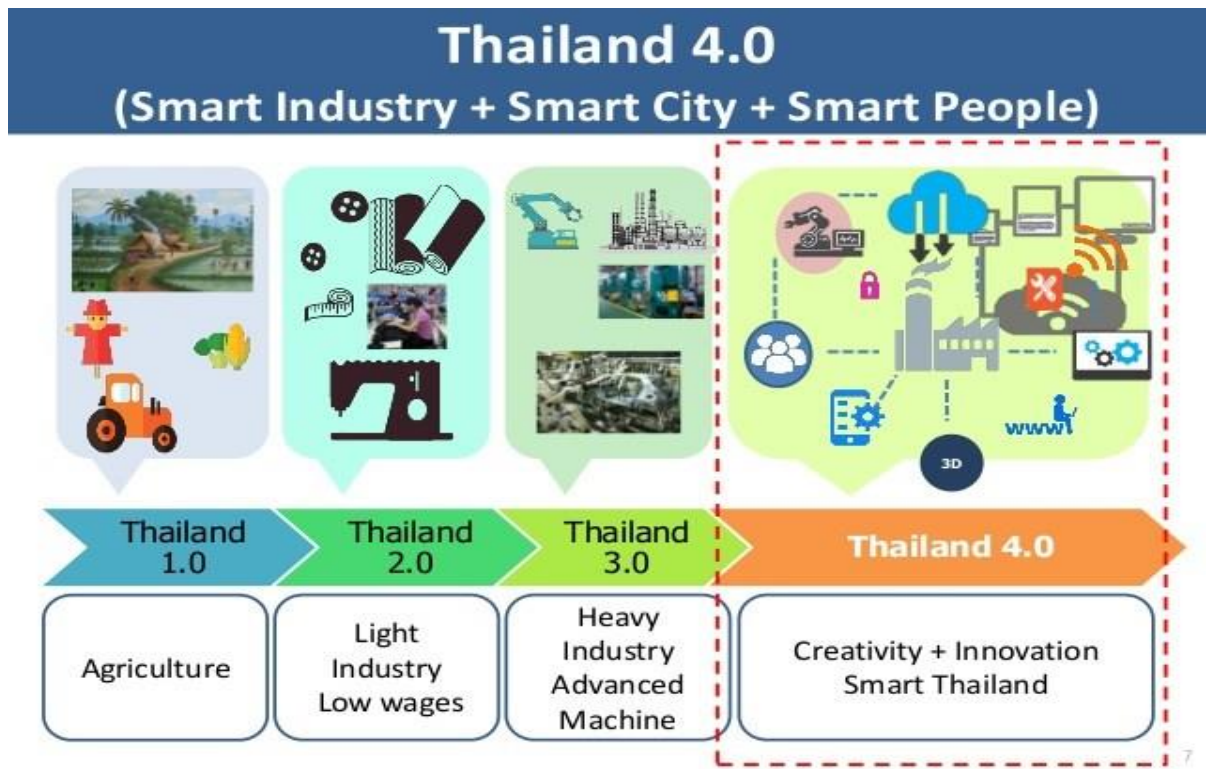

Figure 6: Languepin, O. (2018, October 4). *Thailand 4.0, What Do You Need To Know?* Retrieved from Thailand Business News: https://www.thailand-business-news.com/economics/54286-thailand-4-0-need-know.html

From Figure 6, Thailand follows its predecessors in its naming and purpose – to show the next sector primed for development. The first stage was focused on agriculture, followed by light industry with low wages. Thailand is currently in transition between Thailand 3.0, which focuses on heavy industries with advanced machinery, and Thailand 4.0 which shifts to an economy that is digitally oriented and innovation driven, focusing on high-value-added manufacturing and services (The Economist, 2017).

Thailand 4.0 lays out the fundamentals to FinTech development, highlighting Thailand's potential to leapfrog early-stage digital banking services and move towards integrated digital commerce and other value-added FinTech services (Leesa-Nguansuk, 2017). And this is needed, with Thailand's digital banking penetration standing at 19%, credit card usage of 3.7% (Leesa-Nguansuk, 2017), smartphone penetration at 30% and 68% of the Thai population using cash as their preferred mode of payment (United Overseas Bank, 2017).

The junta has selected 10 strategic industries for development, of which FinTech initiatives do not just serve as an enabler but also forms one of these 10 industries, namely digital services. The first 5 S-Curve are industries that are currently found in Thailand (e.g. Electrical and Electronics, Automotive, Tourism, Jewellery and Gems and Agriculture) where value can be added with new technologies. The 5 new S-Curve will focus on developing industries that are new to the country (The Economist, 2017). Those new industries include Robotics, Aviation & Logistis, Biofuels & Biochemical, Digital and Comprehensive Medical Industry. An explanation offered by Accenture about S-Curve fits with Thailand 4.0 policy.

Its argument is that market growth will stop at some point in time, and it becomes a necessity to migrate to the next opportunity. This is known then as "jumping" to the next S-curve (Nunes & Breene, 2011). However, this is ambitious. Thailand lacks the domestic capabilities to develop these 10 industries, hence the success of the policy will be heavily dependent on Thailand's ability to attract foreign direct investment (FDI). New legislations have come into force that seeks to provide incentives, which includes tax exemption, expanding shareholding options for foreigners, to name a few (The Economist, 2017). A recent report from United Nations show that FDI to Thailand rose by 3.7 times due to increased inflows from European Union (EU) and ASEAN. However, a look at FDI inflows data from 2012 (9135 millions) to 2017 (7635 millions) shows that Thailand has been struggling from below-par FDI figures. This negative trend is made starker when compared with FDI inflows for the entire region of South-East Asia and against ASEAN members. The Economist Intelligence Unit argues that the negative trend is caused by 1) scarcity of skilled labour and 2) weakening domestic market for goods and services. For point 1, UNESCO stated in its Global Education Monitoring Report in October 2017 that only 50% and 46% of students had a minimum proficiency in reading and mathematics respectively, upon completion of lower secondary education. These statistics have remained the same from 2003 to 2015, showing no improvements. For point 2, private consumption growth only averaged 2.8% per year from 2012-2016. In comparison, Indonesia and Malaysia posed 5.2% and 6.9% respectively within the same period. A sluggish growth in private consumption shows fragile consumer confidence in Thailand's economic prospect and one cause of this is the political uncertainty that still surrounds the country (The Economist, 2017).

This analysis shows the challenges that Thailand needs to overcome to increase investors' confidence in its economy, ultimately securing the success of Thailand 4.0. At the same time, this sets out the context for analysing some initiatives that are proposed and or implemented by Thailand, more specifically to encourage the development of FinTech.

### 4.3.2 Regulatory Sandbox

Like the findings of Singapore and Malaysia, Thailand has its version of a regulatory sandbox, introduced by the Bank of Thailand (BOT) on 21 December 2016. BOT has identified FinTech to be a vital factor in developing the financial services industry and 1 of the 10 S-Curve sectors – Digital Economy in the Thailand 4.0 policy. To this end, BOT seeks to encourage the development of FinTech through (1) establishment of a Regulatory Sandbox and (2) support for FinTech Start-ups by introducing financial institutions to Start-ups through a platform, increasing chances of partnership (Corbett, 2017). The sandbox specifies the qualifications for entry and participation into the sandbox and the rules that participants must follow once granted entry into the sandbox.

BOT has outlined its goals of utilising a regulatory sandbox, which are:

- To promote the innovation's development which will increase choices for customers, improve access to financial services and decrease cost
- To promote FinTech testing in a limited environment which would decrease time of implementation into live markets and increase scope of service into wider areas
- To limit risk to the overall financial system
- To promote learning and exchange of information between sandbox participant and BOT, leading to proper governance and preventing impediments to the innovation's development

BOT, comparatively with Singapore and Malaysia, has laid out an exclusive guideline on its target audience, which includes financial institutions, FinTech firms and general technology firm (Bank of Thailand, 2016). Moreover, applicants are expected to have measures proving good corporate governance, data protection and confidentiality, security of work processes and information, AML/CFT and proven track record in managing assets. BOT has also scoped the types of innovation that can be tested in its sandbox, namely lending, payment and transactions deemed appropriate by BOT – essentially a black box. Additionally, participating in the sandbox does not exempt the company from applying for other licenses necessary for its innovation to operate in Thailand.

There are additional points of contention:

- Financial products and services seeking to enter the Regulatory Sandbox must be a new innovation
- Applicant must present research that shows the potential and risk of the innovation
- Applicant must specify the test's duration, scope and expected results
- Applicant must have sufficient capital to carry out the test and implement the innovation upon exit from the sandbox
- Applicant must have an exit strategy if the test is unsuccessful or an implementation plan if the test is successful

Within the sandbox application, when an applicant has qualified to participate in the sandbox, it must conduct the following:

- Provide consumers with accurate, timely information during the testing period
- Consumers have complete control in selecting their choice of financial products/ services

- Provide channels to consumers to receive feedback and manage complaints
- Allow monetary claims to consumer should they suffer financial loss, not caused by them, during the testing period
- Must have safeguards against AML/ CFT
- Must adhere to related laws
- Timely submission of test reports that shows statistical and risk information, key performance indicators, system errors and consumer complaints (if any) to BOT, during and after the testing period

The need to comply to these qualifications before applying for the Regulatory Sandbox could become major obstacles to foreign companies and smaller FinTech Start-ups, which leads to questions over the effectiveness of the Sandbox. Regardless, it is a relatively new initiative by BOT, so it remains to be seen how successful it is in achieving its objectives. Moreover, applicants must prove that their innovation and or technology is new, which is defined by being different from existing products or financial services in Thailand. This is similar to a requirement in Malaysia's Regulatory Sandbox and this could inhibit the element of innovation. An important consideration is whether improvements to existing products or financial services can be allowed to participate in the sandbox for it may fulfil BOT's objective of decreasing cost of business, increase choices for consumers and promote the development of innovations.

### 4.3.3 National e-Payment Master Plan

In October 2016, BOT launched its National e-Payment Master plan that is aimed at promoting the use of e-Payment across all sectors. With this, BOT started with an electronic money transfer service called PromptPay (Chantanusornsiri & Banchongduang, 2016). This project would see the development of an infrastructure that would help 3 wide sectors, namely public, private and the general public to achieve two aims, one to make it more convenient and cheaper to transfer funds and support e-Commerce growth (Bank of Thailand, 2016). In addition, BOT pushed forward a second project – Card Usage Expansion Project which seeks to expand Electronic Data Capture (EDC) distribution and promote the use of debit cards instead of cash. This aims to decrease costs associated with cash management and allow both businesses and the general public to experience the convenience of using card-based payments over cash.

Using PromptPay allows transfers to be carried out conveniently. There is no longer a need to obtain the receiver's bank account number, but instead transfers can be done with national ID card number, corporate tax ID number or mobile phone number. Moreover, the service fees associated with PromptPay is lower than traditional transfer services, as seen in table 2. With these rates, it would incentivize a switch to PromptPay and also to increase the value and volume of transactions. However, PromptPay brings bad news for commercial banks. The lower service fees will cut into banks' revenue. For example, Kasikorn's money transfer fees make up 2% of the bank's total revenue, and PromptPay would see its money transfer fees being a tenth of what it previously was. For now, PromptPay is for users who are banked, but there are future plans to introduce an e-Wallet service to PromptPay and it remains to be seen if users who are unbanked would be able to use the service.

Considering the emerging trend of the population using mobile banking for fund transfers, increasing from 27.4% to 43.8% in the 2015-2016 period, the development of a common infrastructure found in PromptPay is welcomed by consumers, allowing the scaling of reach to remote/ rural parts of Thailand. Over the course of 2015-2016, Thailand has seen a 90.2% growth in the value and a 123.6% growth in volume, transferred using mobile banking.

Using e-Payment is not without its risk. As such, BOT is focused on supervising service providers to ensure that there are appropriate security and alignment with robust standards to build confidence in using e-Payments. BOT established the Guiding Principles for Trusted Mobile Payments. It serves as a guideline to build and enhance good service standards so that users are confident and increase their usage of mobile payments. The guideline is made up of 6 key principles:

1) Risk Management: Providers must have risk management processes that cover IT security, cyber and operational risk
2) Secure Authentication: Providers must have process and methods to protect against identity theft
3) Consumer protection and education: Protect users against risks or potential frauds, including unauthorized access to users' confidential data. To education users to grow their awareness in payment transaction security
4) Openness and Interoperability: Support fair competition and serve innovative choices for customers. Promote investments in IT infrastructure that supports efficient mobile payments
5) AML/CFT and fraud protection: Providers must comply with relevant laws and processes such as customer identity verification and Know Your Customer/ Customer Due Diligence (KYC/CDD)
6) User experience: Create satisfaction of user experience which will change users' behaviour to use their mobile for payments.

Another initiative is to increase collaboration between relevant regulators to share information and expertise, resulting in consistent regulatory frameworks. BOT acknowledged that technological advancement has led to a rapid development and unprecedented changes in the financial and payment services. These pose challenges to regulators as they are exposed to new types of risks and the changing nature of the scope of users. BOT has signed a Memorandum of Understanding (MOU) with the office of The National Broadcasting and Telecommunications Commission (NBTC) to conduct a study in a narrow, but important scope on regulating payment services used via mobile and wireless devices. This strategic relationship is important. With PromptPay as an example, changing SIM card is no longer a straightforward exercise as PromptPay accounts are linked to mobile phone numbers. As such, NBTC has set new standards in authenticating users for new SIM card issuances, change in SIM card ownership and personal information change requests. The new standards require users to bring their national ID cards when carrying out these services which will enhance financial transaction service security and prevent against making transfers to incorrect mobile numbers or mobile numbers assumed by criminals for financial fraud.

# 5. Recommendations

The importance and necessity of mobile financial services in Myanmar is unquestionable. Yet, this industry is still in a nascent stage. As seen in the findings of this report, the role of governments in encouraging and developing economic sectors are crucial, and this is done through policies highlighting the governments' priorities and strategies to achieve success. To this end, the Myanmar government presented an economic policy on July 2016 highlighted the government's push to focus on 12 priorities (Kyaw & Hammond, 2016). In these 12, there are 6 of them that are relevant to this report, which are:

1. Supporting competition and the private sector by practicing a market-oriented system in every sector, cut unnecessary red tape, dilute the power of monopolies and expand access to credit
2. Infrastructure development which will have one of its focus areas on producing and distributing power
3. Support agricultural sectors to encourage inclusive growth, boost food security, increase exports and enhance living standards. One of the ways to achieve this is to increase access to credit for farmers
4. Welcoming FDI by promoting responsible business and creating a stable environment through improving property rights and enforce rule of law. In all, companies will be convinced that Myanmar is secure to invest in
5. **Creating a financial system that can provide capital to businesses, farmers and households in a sustainable manner. The government acknowledges that the financial sector is underdeveloped and excludes large sectors of the economy. The government will review limitations on bank lending and enable mobile banking among other initiatives**
6. **Helping SMEs by improving the ease of doing business in Myanmar, increase access to financial services and develop a skilled workforce**

It is apparent that FinTech, especially mobile financial services, would become an important enabler for the government to achieve the goals set out in its economic policy. Hence, the following recommendations will be based off the points listed above and findings of neighbouring countries' policies, strategies and best practices. However, as a disclaimer, these are my personal recommendations which are heavily derived by studying the examples of Singapore, Malaysia and Thailand. There are many political, societal and economic factors that are not privy to me, therefore the Myanmar government could or are already heading in a different direction pertaining to the area of MFS development.

## 5.1 Adoption of Mobile Financial Services as a key focus area

The countries analyzed in this report have specifically identified FinTech as an area to develop and support. In the same way, it would serve Myanmar well to do the same. However, I opine that the focus should be taken up by Central Bank Myanmar (CBM) and not the wider Myanmar government so that the adoption process can be sped up and not be subjected to long discussions in the house.

In Singapore, Malaysia and Thailand, taskforces have been created to spearhead growth and new initiatives in FinTech. This would allow expediency and lend focus in developing the industry.

Likewise, it would be wise for Myanmar to do so, creating a taskforce that is housed in CBM and it is crucial for CBM to empower this taskforce to make decisions in mobile financial services.

Myanmar's financial system is subject to distrust by the population, with the banking crisis in 2003, lengthy processes in opening accounts and carrying out basic banking transactions and the lack of basic banking infrastructure contributing to 1) 80% of people being outside of the formal financial system and 2) a greater need for mobile financial services (Minischetti & Gallery, 2018). To tackle this, consumer education and community engagement are crucial to increase the level of trust and convince users to utilize MFS.

## 5.2	Modified payment infrastructure

Following the examples of PromptPay, PayNow and PayNet in Thailand, Singapore and Malaysia respectively, it would be beneficial for Myanmar to have a shared payment infrastructure where the public and private sector can utilize. Not only that, but this will greatly increase access to financial resources to every part of Myanmar as long as the user has an Internet-enabled mobile phone.

But instead of creating one, I believe using one of the existing players on the market would be more prudent as it does not require large capital outlay from the government and efficient in getting to market. This would be like the approach taken by Malaysia. They chose PayNet, an existing market player, to be the shared payment infrastructure in the country.

This would also be an initiative of interest in part of addressing Myanmar's issue with tax takes, and this would segue into the next recommendation of encouraging FDI inflows.

## 5.3	Encourage FDI inflows

Myanmar has one of the lowest tax takes in the world, amounting to just 6.4% of GDP. This highlights the lack of domestic capacity to push any tangible reform (The Economist, 2018). UNCTAD issued a World Investment Report in 2018 and categorized Myanmar as a structurally weak and vulnerable economy. Despite FDI inflows increasing by 45.2% year-on-year to US$4.3bn in 2017, this still represents a low absolute number. My view is that the usage of MFS, especially a shared payment infrastructure would increase the tax takes and address the difficulty of tax collection in a significant way. However, another issue that requires discussion is shoring investors' confidence.

It is important for the Myanmar government to cut unnecessary red tape and onerous requirements to encourage starting of new businesses. Following Thailand's 4.0 policy, Myanmar could exercise flexibility and offer incentives for foreign corporations to invest into local companies that meets its economic reform policy. In its new Myanmar Companies Law (MCL), foreign investors can own up to 35% of a local company and still maintain its local company status. But 35% still represents a low equity percentage for investors. Also, Myanmar can follow in Thailand's steps to offer special Visas for investors, executives and individuals who are involved in crucial industries, especially in FinTech to live in Myanmar. The caveat for this to work is that there must be cross-ministry collaboration to share information and achieve the bigger picture. This special Visas would allow for skilled labour to not only come in and develop economically-significant industries, but also to share knowledge and develop the local workforce – all of importance to set the foundation for a more robust economy. To some extent, the Myanmar government has started granting visa-exemptions to some countries, with most coming from ASEAN. In my view, this is the low-hanging fruit which Myanmar should aim to provide visa-free status to all ASEAN member countries and extend it to other nationalities.

### 5.4  Modified Regulatory sandbox

FinTech is set to take the world by storm. Governments need to be prepared to synergise these disruptions with the prevailing financial systems and regulatory sandboxes are an excellent way to do so. Southeast Asian governments should be quick to design and operate sandboxes or risk losing out in the long term (Gnanasagaran, 2018). Across the 3 countries that were explored in this report, every regulatory sandbox has its nuances in qualifying conditions, operations and purpose.

Myanmar should introduce a modified regulatory sandbox that would not only allow experiment on innovative FinTech products/ services/ solutions, but also to include networking which will allow non-bank companies and startups to partner up with banks to experiment and potentially launch in partnership. The latter objective is to ensure that upon a successful exit from the sandbox, the FinTech innovation will be collaborative and not disruptive which in my opinion would be an easier, low-hanging fruit to achieve and sustain over disruptive FinTech innovations.

In the qualifying conditions, it should follow Singapore's model, one that is as open as possible with the conditions to encourage more businesses, bank and non-bank to enter the sandbox. The reason for this is that volume would help the Myanmar government to experience the rapid progress of FinTech, learn and adjust current legislations to account for these innovations. However, some guidelines should still be present which includes AML/CFT prohibitions, proper risk management process and a robust data protection and confidentiality plan.

In the operation of the sandbox, Myanmar should follow its counterparts in relaxing current laws and other red tape. The aim of it for the regulator is to see the potential and benefits of FinTech innovations on the economy, which would define the regulator's role to be one that does not stifle but encourages creativity and innovation.

With a successful exit from the sandbox, regulators should support the FinTech innovation's deployment into the economy. Support can come in the form of expedited application for necessary licenses and or allowing partnerships with banks that have the existing licenses and adding on additional modules in these licenses. The modularity approach would increase efficiency by reducing redundant workload for both the government and company.

## 6.  Summary

In summary, there are 4 main recommendations that the Myanmar government could consider implementing. They are:

1) Adopting Mobile Financial Services as a focus area
    - Create a taskforce out of CBM to spearhead initiatives to develop the MFS industry
    - Tackle the citizenry's general distrust in the country's financial system through community engagements and user education
    - Allocation of resources to develop MFS as a focus area
2) Introduce a shared payment infrastructure

- Follows PayNow (Singapore), PromptPay (Thailand) and PayNet (Malaysia)
- Addresses the current issue of fragmentation for payment/ fund transfer players in the market
- Should not attempt to build a shared payment infrastructure from scratch – PromptPay model, but leverage an existing player in the market and establish an exclusive partnership – PayNet model
- To be used by public and private sector, and it will increase access to financial resources for all
- Can help to increase tax takes for the government

3) Encourage FDI inflows
    - Myanmar lacks the domestic capacity due to low tax takes to develop the MFS industry
    - Need FDI to support sectoral development
    - The government needs to shore up investors' confidence to make Myanmar an attractive proposition for investment.
        - Cut unnecessary red tape and increase the ease and efficiency of conducting business
        - Offering visa exemptions or special visas to investors, executives or individuals who are involved in crucial industries, which in this case, FinTech
        - Consider allowing higher equity ownership in local companies for foreign investors

4) Introduce a Regulatory Sandbox
    - Allow experimentation in a test environment to assess the potential benefits and risk of the innovation. During the operation of the sandbox, regulators will determine the appropriate safeguards to manage risk but not stifle innovation and growth
    - Regulators are to be catalysts in matching innovative solutions to banks or MNCs to bring to market upon exit from the sandbox. This ensures that proposed solutions will be collaborative and not disruptive, benefitting users and the economy
    - Relax certain regulations during the operation of the sandbox
    - Regulators should support go to market opportunities by expediting license applications or even providing waivers. Also, regulators should encourage partnership with banks and exercise flexibility by allowing Startups or FinTech companies to be covered by the banks' licenses

Regulators should consider introducing a modular-based framework, which allows licenses to be granted and modified by adding on modules, as opposed to applying for a new license whenever a new product or service is introduced